\documentclass[prb,preprint]{revtex4-2} 

\usepackage{amsmath}  % needed for \tfrac, \bmatrix, etc.
\usepackage{amsfonts} % needed for bold Greek, Fraktur, and blackboard bold
\usepackage{graphicx} % needed for figures
\usepackage{color}

\begin{document}

\title{Explicit Reconstruction of Polarization Ellipse \\ 
using Rotating Polarizer}

\author{Ramonika Sengupta}
\email{ramonika.sengupta@gmail.com} 
\author{Brijesh Tripathi} 
\email{	Brijesh.Tripathi@sse.pdpu.ac.in}
\affiliation{Pandit Deendayal Petroleum University, Raisan, Gandhinagar 382421  INDIA}

\author{Asha Adhiya}
\email{ashaadhiya.ipr@gmail.com}
\affiliation{Institute for Plasma Research, Bhat, Gandhinagar-382428 INDIA}

\begin{abstract}
This paper describes a method for the explicit reconstruction and visualization of various polarization ellipses from the intensities measured after a rotating polarizer. The state of polarization of a light beam is represented by the variation of the electric field amplitude vector with polar angle in the laboratory coordinate system defined by the experimental set-up. The intensity of the light beam measured after passing through a rotating polarizer gives the estimate of the wave electric field component parallel to the polarizer pass axis averaged over a time period. The formulae for the estimation of orthogonal field component and polar angle, in the polarizer’s coordinate system have been derived. Both the orthogonally polarized components and the polar angle are functions of ellipse parameters that have been deduced from the intensity measurements. Finally, the resultant electric field and the polar angle are mapped to the laboratory coordinate system and compared with the representation of the polarization ellipse obtained from the Stokes vectors.
\end{abstract}

\maketitle % title page is now complete

\newpage
\section{Introduction} % Section titles are automatically converted to all-caps.
% Section numbering is automatic.
Electromagnetic waves are represented in terms of their intensity, wavelength, and state of polarization. The general form of a state of polarization is an elliptical polarization, with linear and circular polarizations being the special cases~\cite{Born},\cite{Goldstein}. 
The state of polarization is completely characterized by the semi-major and semi-minor axes, orientation, and ellipticity of the polarization ellipse. The axes, orientation, and magnitude of ellipticity can be directly obtained by measurements using a simple optical component like a polarizer. The interpretation of polarization experiments generally employs Stokes polarization parameters \cite{Jones},\cite{Schaefer} because: (i) the polarization ellipse, which is an amplitude description, is not directly accessible to measurement; and (ii) the parameters of the polarization ellipse can be deduced very easily from Stokes parameters. It would be interesting to have a point-by-point correspondence between the experimentally measured intensities and the polarization ellipse as expected from theory. However, as mentioned in the reference \cite{Fernando}, such correspondence is not available in the standard textbooks.

The main issue in interpreting the polarization experiments is that the polarization state is expressed in terms of the radius vector of the polarization ellipse, which is traced out on time scales much faster ($\sim 10^{-15}$ s for visible light) than the measurement time scales. On the other hand, the measurement in polarization experiments is in terms of intensity, which is proportional to the time average of the square of the radius vector of the ellipse. Further, with polarization optics inserted in the beam path, the measured intensity gives the estimate of the time average of the square of only one component of the radius vector (viz., along the pass axis of the polarizer). In earlier works, the intensities of known elliptical polarizations have been theoretically estimated~\cite{Mayes},\cite{Cox}. %The experiments have also been carried out to promote students' understanding of polarization~\cite{Anna}. 
Graphical methods to reconstruct the polarization ellipse from the measured intensities after polarizing optics have also been demonstrated~\cite{Fernando}. 
To the best of our knowledge, there is no study available that gives an analytical reconstruction of the polarization ellipse from the measured intensities. This paper describes a method to explicitly reconstruct the polarization ellipse from the intensity measured as a function of the angle of rotation of a polarizer. The corresponding experimental uncertainties have also been estimated.

The paper is organized as follows: Section~\ref{sec:ell_pol_light} gives the general description of elliptically polarized light and the representation of the polarization ellipse in the lab frame. The method of reconstruction of the polarization ellipse from the measured intensities is given in Section~\ref{sec:reconstruct}. The experimental setup is described in Section~\ref{Expt}, with the procedure for reconstruction of an ellipse summarized in Section~\ref{Summary}. The experimental results and discussions are given in Section~\ref{results}, and the conclusions presented in Section~\ref{conclusion}.

\section{ELLIPTICALLY POLARIZED LIGHT}
\label{sec:ell_pol_light}

\begin{figure}
\centering
\includegraphics[width=0.55\textwidth]{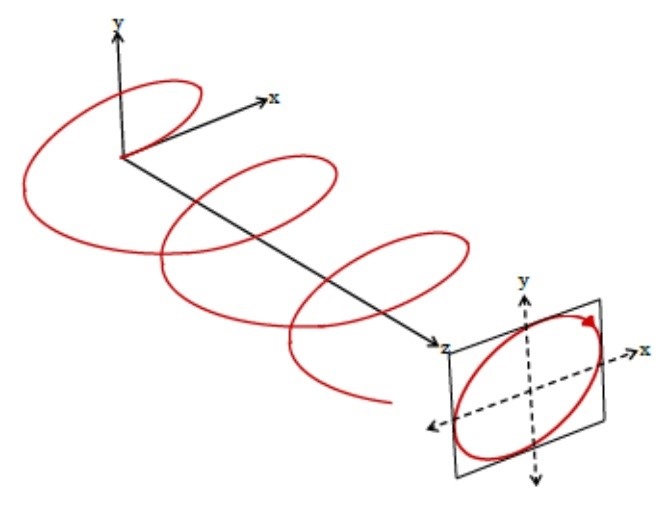}
\caption{Electric field vector $\vec{E}$ in space at a particular instant of
time for an elliptically polarized light (rep) and the ellipse traversed by the
vector in the $xy$-plane.}
\label{Figure1}
\end{figure}

Elliptical polarization is the most general form of the state of polarization. For the light propagating along the $z$-axis, it can be described as the superposition of two orthogonal linearly polarized components (see Fig.~\ref{Figure1}):  
\begin{subequations}
\begin{align}
\Vec{E}  & =  E_x(z,t) \hat{x} + E_y(z,t) \hat{y} \\
 & =  E_{ox} \cos(\omega t -kz + \delta_x)\hat{x} + E_{oy}\cos(\omega t -kz + \delta_y)\hat{y}
 \label{Eqn1}
\end{align}
\end{subequations}
where $E_{ox}$ and $E_{oy}$ are the maximum amplitudes of electric field components $E_x(z,t)$ and $E_y(z,t)$ along $x$-axis and $y$-axis, $\delta_x$ and $\delta_y$ are the initial phase angles of $E_x(z,t)$ and $E_y(z,t)$, $\omega ( = 2 \pi \nu)$ is the angular frequency, $\nu$ is the frequency of the electromagnetic wave, $\Vec{k} (=~2\pi/\lambda~\hat{z})$ is the wave vector along $z$-axis, and $\lambda (= c/\nu)$ is the wavelength of the electromagnetic wave. 

The tip of the electric field vector $\Vec{E}$ describes an ellipse in the $xy$-plane when viewed from the positive $z$-axis as shown in Fig.~\ref{Figure1}. The equation of this ellipse in the $xy$-plane is obtained by eliminating the space-time propagator $\tau = \omega t - kz$ (see e.g., \cite{Goldstein}, Eq.~3-7):
\begin{eqnarray}
\frac{{E_x}^2}{{E_{ox}}^2} + \frac{{E_y}^2}{{E_{oy}}^2} -\frac{2E_x E_y}{E_{ox}E_{oy}} \cos \delta & = & \sin^2\delta
\label{Eqn2}
\end{eqnarray}
where $\delta = \delta_y - \delta_x$ is the phase of the $y$ component
relative to the $x$ component of the electric field.
\begin{figure}
\centering
\includegraphics[width=1.0\textwidth]{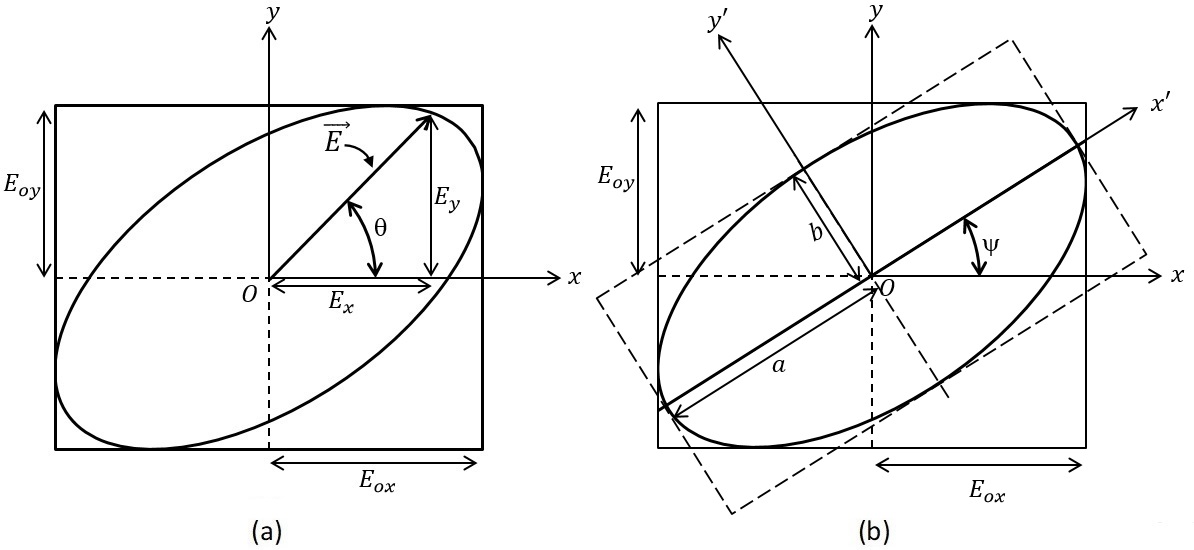}
\caption{Rotated polarization ellipse (Eq.~\ref{Eqn2}) in $xy$-coordinates, i.e., lab frame showing: (a) $\vec{E}$ at a particular instant of time; (b) Standard form of ellipse~(Eq.~\ref{Eqn3}) in $x'y'$-coordinates.}
\label{Figure2}
\end{figure}

Eq. (\ref{Eqn2}) describes the rotated ellipse, i.e., in its non-standard form as shown in Fig.~\ref{Figure2}(a) with $\Vec{E}$ as the radius vector. At a given instant of time, $\theta$ is the angle between $\Vec{E}$ and $x$-axis such that:
\begin{eqnarray}
E_y = E_x\tan \theta
\label{Eqn9}
\end{eqnarray}
We can write the equation of an ellipse in standard form in the $x'y'$-coordinates (see Fig.~\ref{Figure2}(b)) that is  rotated by an angle $\psi$ with respect to $xy$-coordinates as (\cite{Goldstein}, Eq. 3-26):  
\begin{eqnarray}
\frac{{E'_x}^2}{a^2} + \frac{{E'_y}^2}{b^2} = 1
\label{Eqn3}
\end{eqnarray}
where $a$ is the major radius of the ellipse along the $x'$-axis and $b$ is its minor radius along the $y'$-axis. The resultant electric field vector $\Vec{E}$ is given as: 
\begin{subequations} \label{Eqn4}
\begin{align}
\vec E  & =  {E'}_x(z,t) \hat{x'} + {E'}_y(z,t) \hat{y'} \\
 & =  a\cos(\omega t -kz + \delta_o)\ \hat{x'} + b\sin(\omega t -kz + \delta_o)\ \hat{y'}
\end{align}
\end{subequations}
where $\delta_o$ is an arbitrary initial phase angle common to both the $x$ and $y$ components. It can be eliminated by an appropriate choice of origin ($z = 0$) and initial instant ($t=0$) without any loss of generality.

Now we want to correlate the electric field components in the lab coordinates (i.e., ${E}_x$, ${E}_y$) to the parameters that can be measured in the experiments.  To do so, we return to the standard form of the ellipse, as given in Eq.~(\ref{Eqn3}), and write ${E'}_x$ and ${E'}_y$ as (\cite{Goldstein}, Eq.~3-24):
\begin{subequations} \label{E_comp}
\begin{align}
{E'}_x & =  {E}_x \cos \psi +  {E}_y \sin \psi \\
{E'}_y & = - {E}_x \sin \psi +  {E}_y \cos \psi  
\end{align}
\end{subequations}
Then, the equation of ellipse becomes:
\begin{eqnarray}
\frac{{[{E}_x \cos \psi +  {E}_y \sin \psi ]}^2}{a^2} + \frac{{[{E}_y \cos \psi -  {E}_x \sin \psi]}^2}{b^2} = 1
\label{Eqn7}
\end{eqnarray}
such that
\begin{multline}
\quad \quad (b^2\cos^2\psi + a^2\sin^2\psi){E_x}^2 + 2(b^2-a^2)\sin\psi \cos\psi E_x E_y \\
+ (b^2\sin^2\psi + a^2\cos^2\psi){E_y}^2 -a^2b^2 = 0 \quad \quad \quad \quad
\label{Eqn8}
\end{multline}
Substituting $E_y$ from Eq.~(\ref{Eqn9}) in Eq.~(\ref{Eqn8}), we get:
\begin{subequations} \label{E_xy_coord}
\begin{align}
E_x  & =  \pm \frac{a b \cos \theta}{\sqrt{\mathcal{D}_1 +\mathcal{D}_2 +\mathcal{D}_3}}\\
E_y  & =  \pm \frac{a b \sin \theta}{\sqrt{\mathcal{D}_1 +\mathcal{D}_2 +\mathcal{D}_3}}\\
|\vec{E}|  & =  \frac{a b }{\sqrt{\mathcal{D}_1 +\mathcal{D}_2 +\mathcal{D}_3}}
 \quad \quad \quad \quad \quad \quad 
\end{align}
\end{subequations}
where the terms in the denominator are given as:
\begin{subequations} \label{Den_terms}
\begin{align}
\mathcal{D}_1  & =  (a^2 \cos^2 \theta + b^2 \sin^2 \theta)\sin^2 \psi \\
\mathcal{D}_2  & =  (b^2 \cos^2 \theta + a^2 \sin^2 \theta)\cos^2 \psi \\
\mathcal{D}_3  & =  2(b^2-a^2)\sin \theta \cos \theta \sin \psi \cos \psi
\end{align}
\end{subequations}

Note that the time duration for $\vec{E}$ to trace out the ellipse is one time period $T$ of the electromagnetic wave, i.e., about $10^{-15}$ s at optical frequencies ($\sim 10^{15}$ Hz). Hence, $\vec{E}$, $E_x$, $E_y$, or $\theta$ are not directly observed or measured in the experiments. We measure the intensities of the orthogonal components, viz., $I_x  \propto \left\langle{E_{x}^2}\right\rangle$ and $I_y  \propto \left\langle{E_{y}^2}\right\rangle$, where $\left\langle \, \, \right\rangle$ represents the time average such that $\left\langle{E_{x}^2}\right\rangle = E_{ox}^2/2$ and $\left\langle{E_{y}^2}\right\rangle = E_{oy}^2/2$. As seen from Fig.~\ref{Figure2}(b), we can determine $E_{ox}$  and $E_{oy}$ from the vertical and horizontal tangents to the ellipse in $xy$-coordinates as:
\begin{subequations}
\begin{align}
E_{ox} = \pm \sqrt{a^2 \cos^2 \psi + b^2 \sin^2 \psi} \\
E_{oy} = \pm \sqrt{a^2 \sin^2 \psi + b^2 \cos^2 \psi}
\end{align}
\end{subequations}

\begin{figure}
\centering
\includegraphics[width=1.0\textwidth]{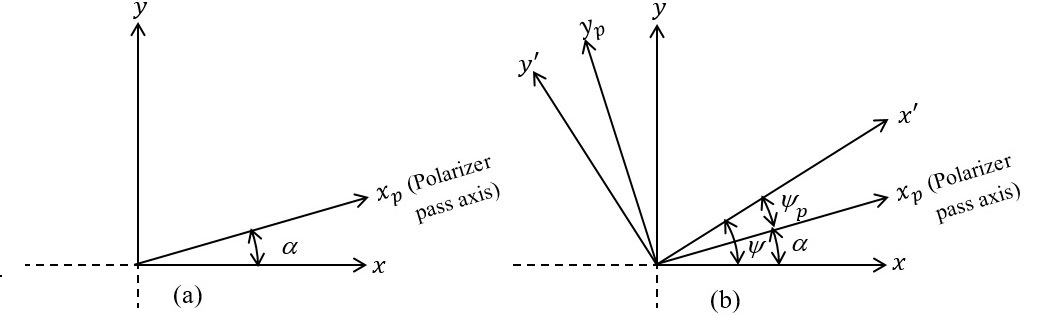}
\caption{Polarizer pass-axis in (a) $xy$-coordinates and (b) $x'y'$-coordinates.}
\label{Figure5}
\end{figure}

In this work, a linear polarizer is rotated in the lab frame such that the angle $\alpha$ of the pass axis of the polarizer changes in the $xy$-plane as shown in Fig.~\ref{Figure5} (a). The resulting intensity  $I_\alpha$ is measured by the detector. 

\begin{figure}
\centering
\includegraphics[width=0.65\textwidth]{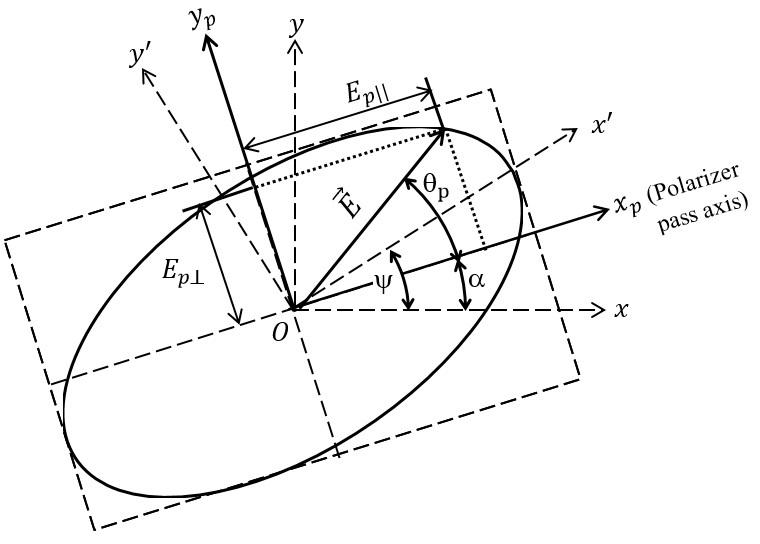}
\caption{Representation of radius vector $\vec{E}$ in polarizer reference frame $(x_p,y_p)$ and its relationship with the lab coordinates $(x,y)$.}
\label{Figure6}
\end{figure}

Now we define a third coordinate system, viz.,  $x_py_p$-coordinates, such that the polarizer pass axis is along the $x_p$-axis. Fig.~\ref{Figure5}(b) shows $\alpha$ in the $x'y'$-coordinates. In addition, we define $\psi_p (= \psi -\alpha)$ as the angle of the major axis of ellipse and $\theta_p (= \theta-\alpha)$ as the polar angle of $\vec E$ in $x_py_p$-coordinates. The normal component $E_{p\perp}$ does not pass through the polarizer, but the electric field component $E_{p||}$ parallel to the $x_p$-axis does (see Fig.~\ref{Figure6}). Analogous to Eq.~(\ref{E_xy_coord}a), we can write $E_{p||}$ as:
\begin{eqnarray}
E_{p||} =  \pm \frac{a b \cos \theta_p}{\sqrt{{D}_1p +\mathcal{D}_2p + \mathcal{D}_3p}}
\end{eqnarray}
where the terms in the denominator $\mathcal{D}_1p, \mathcal{D}_2p $ and $\mathcal{D}_3p$ are obtained by substituting $\theta = \theta_p$ in Eqs.~(\ref{Den_terms}).
\section{RECONSTRUCTION OF POLARIZATION ELLIPSE FROM MEASURED INTENSITY}
\label{sec:reconstruct}	% You can label sections for reference
In this section, we reconstruct the polarization ellipse from the intensity of light measured after passing through the rotating polarizer. With polarizer at angle $\alpha$, we measure the intensity as $I_{\alpha} \propto {\left\langle{E_{p||}^2}\right\rangle}$, where  ${\left\langle{E_{p||}^2}\right\rangle} = E_{op||}^2/2$ such that:
\begin{eqnarray} 
I_{\alpha} \propto |{E_{op||}^2}| \label{Eqn 14}
\end{eqnarray}
\begin{figure}
\centering
\includegraphics[width=0.67\textwidth]{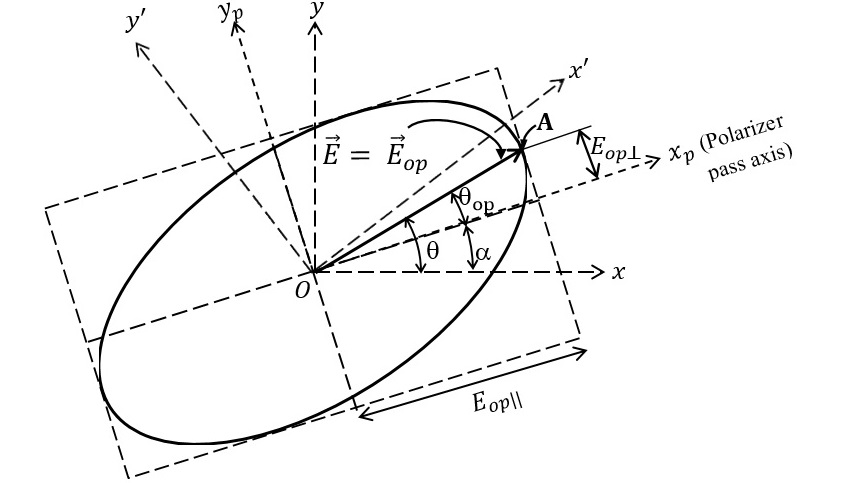}
\caption{Representation of radius vector $\vec{E} (=  \vec E_{op})$, components $E_{op||}$, $E_{op\perp}$, and polar angle $\theta_{op}$.}
\label{Figure7}
\end{figure}
Fig.~\ref{Figure7} shows that $E_{op||}$ gives the horizontal coordinates of the tangent point $A$ on the ellipse in $x_py_p$-coordinates given as:
\begin{eqnarray}
E_{op||}  & = & \pm \sqrt{a^2 \cos^2 \psi_p + b^2 \sin^2 \psi_p}
\label{E_op_par}
\end{eqnarray}
The corresponding vertical coordinates are:
\begin{eqnarray}
E_{op\perp} = \frac{(a^2-b^2)\sin \psi_p \cos \psi_p}{\sqrt{a^2 \cos^2 \psi_p + b^2 \sin^2 \psi_p}} \label{Eqn:15}
\label{E_op_perp}
\end{eqnarray}
Using Eqs~(\ref{E_op_par}) and (\ref{E_op_perp}), the magnitude of the resultant radius vector $\vec E_{op}$ is: 
\begin{eqnarray}
|\vec E_{op}| &=& \sqrt\frac{a^4 \cos^2 \psi_p + b^4 \sin^2 \psi_p}{a^2 \cos^2 \psi_p + b^2 \sin^2 \psi_p} \label{Eqn:17}
\end{eqnarray}
The corresponding polar angle $\theta_{op}$ is:
\begin{eqnarray}
\theta_{op} &=& \tan^{-1} \left[\frac{(a^2-b^2)\sin \psi_p \cos \psi_p}{a^2 \cos^2 \psi_p + b^2 \sin^2 \psi_p}\right]
\label{Eqn:16}
\end{eqnarray}
Therefore, the electric field vector $\vec{E}$ in the lab frame is (see Fig.~\ref{Figure7}):
\begin{eqnarray}
|\vec{E}| =  |\vec E_{op}| \quad \quad  \mbox{ at } \quad \theta = \alpha + \theta_{op}
\label{Eqn:19}
\end{eqnarray}
We reconstruct the polarization ellipse using Eqs.~(\ref{Eqn:17}-\ref{Eqn:19}) from the measured intensity $I_\alpha$ as a function of $\alpha$, as described in Section~\ref{Summary}. 

\section{Experimental Setup}
\label{Expt}

\begin{figure}
\centering
\includegraphics[width=0.86\textwidth]{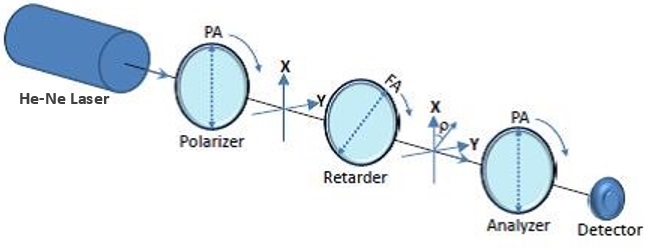}
\caption{Schematic of the experimental set-up.}
\label{Figure8}
\end{figure}

The schematic diagram of the experimental set-up is shown in Fig.~\ref{Figure8}. A 5 mW He-Ne (632.8 nm) laser is used as a source of radiation. The polarizer and analyzer are used to generate a linearly polarized beam and analyze the state of polarization of the outgoing beam, respectively. A quarter wave plate is placed after the polarizer to generate different states of polarization by rotating its fast axis with respect to the plane of polarization of the incident beam. The transmission (i.e., pass axis) and extinction axes of the polarizer are defined as $x$- and $y$-coordinate axes respectively, with the $z$-axis opposite to the direction of beam propagation. The angle $\rho$ is defined as the angle of rotation of the fast axis of the wave plate with respect to the x-axis.  

By changing the orientation angle $\rho$ with respect to the incoming linear polarization, the quarter wave plate retarder generates various states of polarization i.e., polarization ellipses with different orientation angles $\psi$ and ellipticity angles $\chi$. For $\rho = 0^o$ or $90^o$, the emerging beam remains linearly polarized. For $\rho = \pm 45^o$, the emerging beam is circularly polarized (rcp or lcp). At all other values of $\rho$, the emerging beam is elliptically polarized. 

All the optical components are mounted in the kinematic rotation mounts with measurement accuracy of $1^o$. The analyzer is rotated about the z-axis through angle $\alpha$ and the output intensity $I_\alpha$ is measured by a photo detector for one complete rotation. 

\section{Procedure for Reconstruction of Ellipse}
\label{Summary}
With these considerations in mind, we now propose the following procedure for reconstructing the polarization ellipse.
\begin{itemize}
\item For $\rho = 0^\circ$ (viz., linear polarization), the maximum intensity $I_0 = I_{\alpha,max}$ is measured.
\item The intensity $I_\alpha$ is measured as a function of $\alpha$ for different values of $\rho$.
\item For all $\rho$ values:
\begin{enumerate}
    \item \label{itm:1} The measured intensity $I_\alpha$ is normalized by $I_0$.
\item \label{itm:2} The major $a$ and minor $b$ radii are estimated from the maximum and minimum intensity as: 
\begin{eqnarray*}
a = \sqrt{\frac{I_{\alpha,max}}{I_0}}; \quad \quad b =\sqrt{\frac{I_{\alpha,min}}{I_0}}
\end{eqnarray*}
\item \label{itm:3} The orientation of ellipse $\psi $ is determined from the orientation of $a$ viz., $\psi = \alpha_{a}$ 
\item \label{itm:4} The orientation of ellipse $\psi_p$ in $x_py_p$-coordinates is calculated from $\psi_p = \psi - \alpha$.
\item \label{itm:5} Using $a$, $b$ and $\psi_p$ from Steps~\ref{itm:2} and~\ref{itm:4}, we obtain:
\begin{enumerate}
    \item \label{itm:a} From Eq.~(\ref{E_op_par}): $E_{op||}$ as a function of $\psi_p$ or $\alpha \, (=  \psi - \psi_p$)  .
    \item \label{itm:b} From Eqs.~(\ref{Eqn:17}) and~(\ref{Eqn:16}): $E_{op}$ and $\theta_{op}$ as a function of $\psi_p$ or $\alpha$.
    \item \label{itm:c} From Eq.~(\ref{Eqn:19}): $|\vec E|$ as a function of $\theta$ or $\alpha \, (= \theta - \theta_{op}$  ).  
\end{enumerate}
\item \label{itm:6} The polar plots are obtained using $E_{op||}^2$ as a function of $\alpha$ from Step~\ref{itm:a}. 
\item \label{itm:7} The polar plots are obtained using $|\vec E|$ as a function of $\alpha$ from Step~\ref{itm:c}. 
\end{enumerate}
\end{itemize}
Upon completing this procedure, the analytical value of $|\vec E|$ is obtained corresponding to each measurement at $\alpha$. Hence, the polarization ellipse is analytically reconstructed with point-by-point correspondence with measurements. This is entirely different from obtaining the polarization ellipse using values of $a$, $b$ and $\psi$.

\section{RESULTS AND DISCUSSION}
\label{results}
\begin{figure}
\centering
\includegraphics[width=0.98\textwidth]{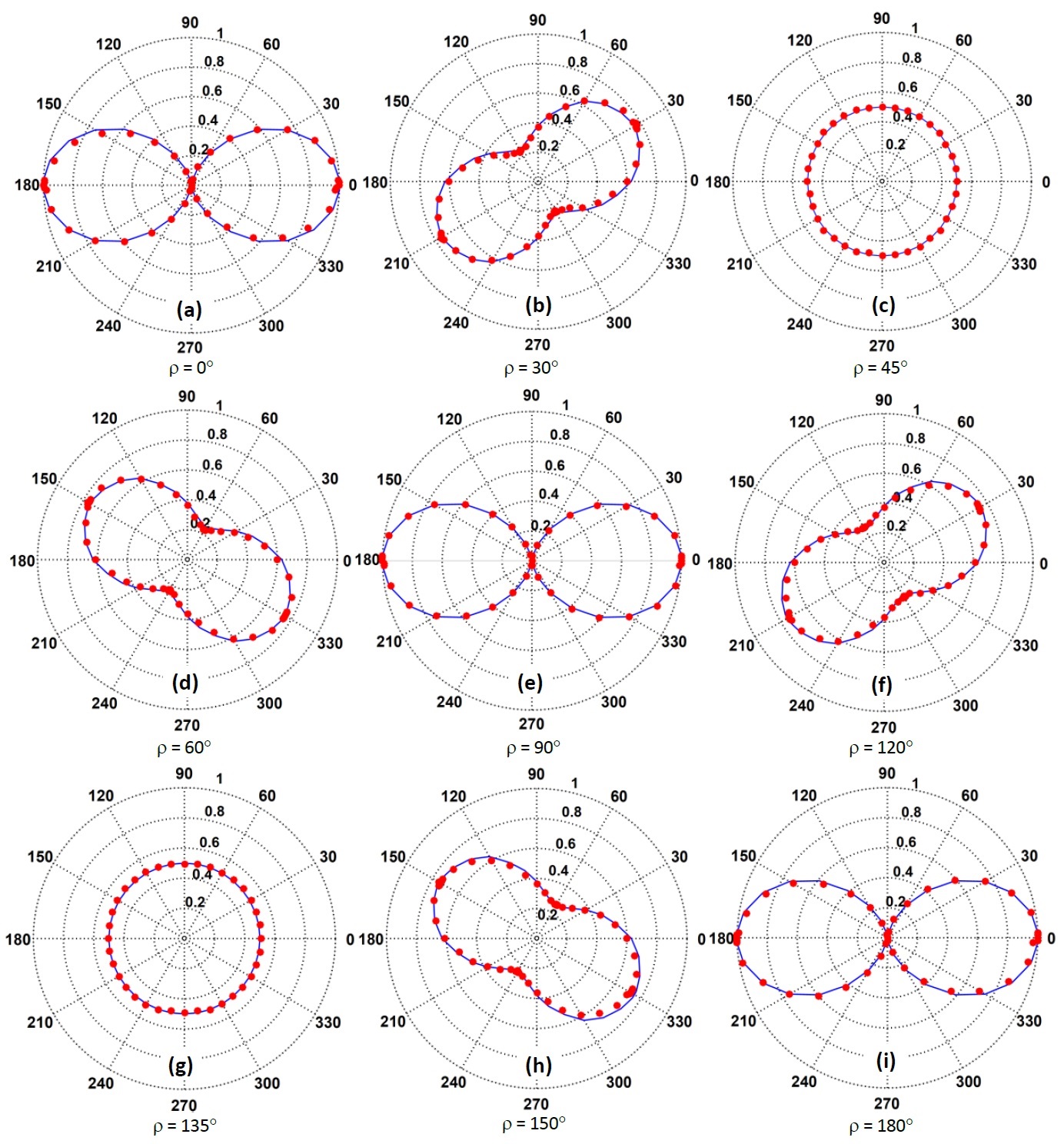}
\caption{Polar plots of normalized $I_\alpha$ ( \color{red}{$\bullet$} \color{black}) 
and ${E_{op||}^2}$ ( \color{blue}{\raisebox{3pt}{\rule{12pt}{0.2pt}}} \color{black}) as a function of angle $\alpha$ for different values of $\rho$, where ${E_{op||}^2}$ is obtained from Eq.~(\ref{E_op_par}), as described in Section~\ref{Summary}, Steps~(\ref{itm:1}) to (\ref{itm:a}). The values of $\rho$ are indicated below each frame.}
\label{Figure9}
\end{figure}

Fig.~\ref{Figure9} shows the polar plots of the normalized $I_\alpha$ as a function of $\alpha$. Frames (a)-(i) show the plots for various values of $\rho$ (mentioned below the frames). The values of $E_{op||}^2$ obtained from Step~\ref{itm:6} in Section~\ref{Summary} are also shown in Fig.~\ref{Figure9}. The values of $I_\alpha$ and ${E_{op||}^2}$ are compared and the maximum difference is found to be less than $0.1$. For $360^\circ$ rotation of $\alpha$, the mean difference between $I_\alpha$ and ${E_{op||}^2}$ is estimated. This mean difference ranges from $0.01$ to $0.04$ for different values of $\rho$ with the maximum observed at $\rho = 60^\circ$. A plausible reason for this difference might be the inaccuracies introduced by the inexact retardation plate used for generating elliptical polarizations. 

\begin{figure}
\centering
\includegraphics[width=0.98\textwidth]{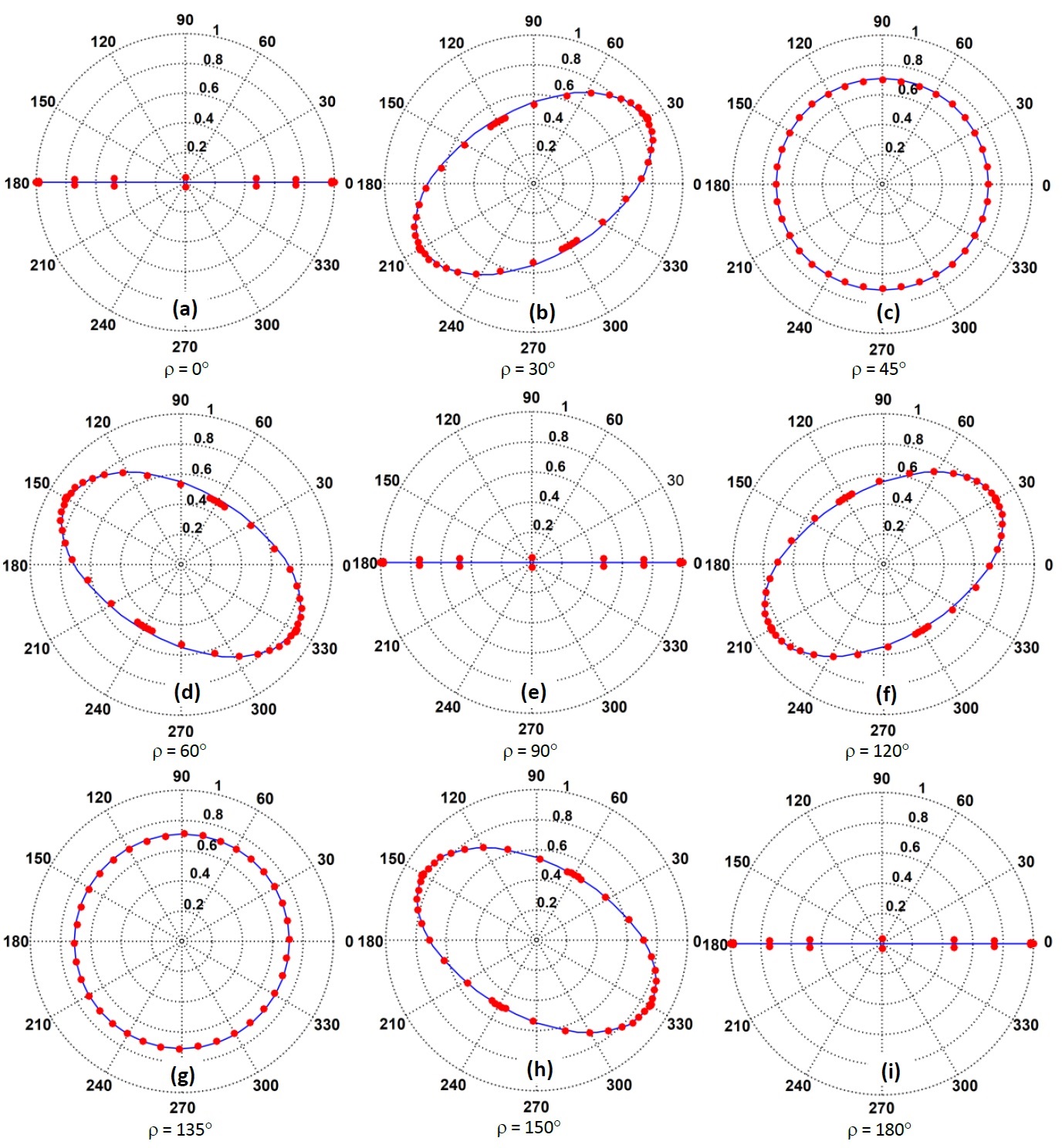}
\caption{Reconstructed polarization ellipse ( \color{red}{$\bullet$} \color{black}) drawn using Eq.~(\ref{Eqn:19}) as described in Section~\ref{Summary}, Steps~(\ref{itm:1}) to (\ref{itm:c}) and the polarization ellipse drawn using parameters $a$, $b$ and $\psi$ ( \color{blue}{\raisebox{3pt}{\rule{12pt}{0.2pt}}} \color{black}) obtained  as described in Section~\ref{Summary}, Steps~(\ref{itm:2}) and (\ref{itm:3}). The values of $\rho$ are indicated below each frame.}
\label{Figure10}
\end{figure}

Fig.~\ref{Figure10} shows the polar plots of $|\vec E|$ obtained from  Step~\ref{itm:7} in Section~\ref{Summary}. As in Fig.~\ref{Figure9}, frames (a)-(i) show the plots for various values of $\rho$ (mentioned on the frames). Fig.~\ref{Figure10} also shows the polarization ellipses drawn using parameters $a$, $b$ and $\psi$ estimated using Steps~\ref{itm:2} and \ref{itm:3} in Section~\ref{Summary}. The difference between the radius vectors obtained by the two methods is less than 0.04. The mean difference the radius vectors over a complete $360^\circ$ rotation of $\alpha$ for various values of angle $\rho$ is less than 0.02 with maximum at $\rho = 60^o$.  

The importance of the present study lies in the fact that it gives the method of recovering the polarization ellipse directly from the experimental measurements. Since the experimental uncertainties are negligible, the measurements give a reliable reconstruction of the ellipse. In addition, the method does not require manipulation of the polarization further by using retardation plates as in the standard methods, the inaccuracies in the reconstruction of the polarization ellipse are reduced.
\section{CONCLUSION}
\label{conclusion}
In this paper, we have described a technique for reconstructing the state of polarization of a light beam using a rotating polarizer. The component of the electric field vector parallel to the polarizer pass axis is directly estimated from the experimentally measured intensity. The formulae to estimate the orthogonal electric field component, magnitude of the resultant electric field vector, and polar angle as a function of the rotation angle of the analyzer in the laboratory coordinates have been derived. These expressions, in turn, are used to reconstruct the polarization ellipse. The intensities from the reconstructed ellipses in the laboratory coordinate system have been estimated and compared with the experimentally measured intensities. The mean difference between the estimated and measured intensities is less than 0.04 for a given polarization. The reconstructed polarization ellipses are plotted. The polarization ellipses obtained from the parameters deduced from the intensity measurements are compared with the reconstructed ellipses. The polarization ellipses obtained by the two methods deviate by less than 0.02 from one another. In conclusion, using a simple apparatus, we have demonstrated a technique for accurately reconstructing the polarization ellipse having a point-by-point correspondence with the experimentally measured intensities.

\end{document}